\documentclass[conference]{IEEEtran}
\IEEEoverridecommandlockouts
\usepackage{cite}
\usepackage[colorlinks=true]{hyperref}
\usepackage{amsmath,amssymb,amsfonts}
\usepackage{algorithm}
\usepackage{algorithmic}
\usepackage{graphicx}
\usepackage{textcomp}
\usepackage{xcolor}
\usepackage{multirow}
\usepackage{subfig}
\def\BibTeX{{\rm B\kern-.05em{\sc i\kern-.025em b}\kern-.08em
    T\kern-.1667em\lower.7ex\hbox{E}\kern-.125emX}}
\begin{document}

\newcommand{\reffig}[1]{Figure~\ref{#1}}
\newcommand{\reftab}[1]{Table~\ref{#1}}
\newcommand{\refsec}[1]{Section~\ref{#1}}
\newcommand{\refequ}[1]{Equation~\ref{#1}}
\newcommand{\etal}{et~al.~}
\newcommand{\ang}{$\text{\AA}$}
\newcommand{\schr}{Schr\"{o}dinger}

\newcommand{\cX}{\mathcal{X}}
\newcommand{\RR}{\mathbb{R}}

\renewcommand{\algorithmicrequire}{\textbf{Input:}}

\title{Towards Accurate Quantum Chemical Calculations on Noisy Quantum Computers}

\author{
    \IEEEauthorblockN{Naoki Iijima\textsuperscript{\textdagger}\textsuperscript{\textsection}, Satoshi Imamura\textsuperscript{\textdagger}\textsuperscript{\textsection}, Mikio Morita\textsuperscript{\textdaggerdbl}\textsuperscript{\textsection}, Sho Takemori*,\\Akihiko Kasagi\textsuperscript{\textdagger}\textsuperscript{\textsection}, Yuhei Umeda\textsuperscript{\textdaggerdbl}, and Eiji Yoshida\textsuperscript{\textdagger}}
    \IEEEauthorblockA{\textit{\textsuperscript{\textdagger}Computing Laboratory, \textsuperscript{\textdaggerdbl}Quantum Laboratory, and *AI Laboratory, Fujitsu Limited}}
    \IEEEauthorblockA{\textit{\textsuperscript{\textsection}RIKEN RQC-FUJITSU Collaboration Center, RIKEN Center for Quantum Computing}}
}

\maketitle
\thispagestyle{plain}
\pagestyle{plain}

\begin{abstract}
Variational quantum eigensolver (VQE) is a hybrid quantum-classical algorithm designed for noisy intermediate-scale quantum (NISQ) computers. It is promising for quantum chemical calculations (QCC) because it can calculate the ground-state energy of a target molecule. Although VQE has a potential to achieve a higher accuracy than classical approximation methods in QCC, it is challenging to achieve it on current NISQ computers due to the significant impact of noises. Density matrix embedding theory (DMET) is a well-known technique to divide a molecule into multiple fragments, which is available to mitigate the noise impact on VQE. However, our preliminary evaluation shows that the naive combination of DMET and VQE does not outperform a gold standard classical method.

In this work, we present three approaches to mitigate the noise impact for the DMET+VQE combination. (1) The size of quantum circuits used by VQE is decreased by reducing the number of bath orbitals which represent  interactions between multiple fragments in DMET. (2) Reduced density matrices (RDMs), which are used to calculate a molecular energy in DMET, are calculated accurately based on expectation values obtained by executing quantum circuits using a noise-less quantum computer simulator. (3) The parameters of a quantum circuit optimized by VQE are refined with mathematical post-processing. The evaluation using a noisy quantum computer simulator shows that our approaches significantly improve the accuracy of the DMET+VQE combination. Moreover, we demonstrate that on a real NISQ device, the DMET+VQE combination applying our three approaches achieves a higher accuracy than the gold standard classical method.
\end{abstract}


\section{Introduction}
\label{sec:introduction}

Quantum computing is currently attracting a lot of attention both in industry and academia, because it is expected to solve large-scale problems that classical computing cannot solve. Quantum computers having tens or hundreds of qubits and noise errors are called noisy intermediate-scale quantum (NISQ) computers, and the recent development of them is remarkable~\cite{IBM_QCRoadmap:2022}. Moreover, a wide variety of variational quantum algorithms have been developed for NISQ computers and studied intensively in some areas such as combinatorial optimization and machine learning~\cite{Cerezo:2021va}.

Variational quantum eigensolver (VQE)~\cite{Peruzzo:2014va} is also one of them and available to quantum chemical calculations (QCC) which are essential for drug discovery and material development~\cite{Klamt:2005co,Bochevarov:2013ja}. VQE calculates the approximate ground-state energy of a target molecule by iteratively executing a parameterized quantum circuit on a quantum computer and updating the parameters of the quantum circuit so that the molecular energy is minimized. VQE is expected to calculate the ground-state energies of large-scale molecules in polynomial time by utilizing the quantum parallelism of NISQ computers. However, its accuracy is degraded significantly by the noise errors of current NISQ computers.

Density matrix embedding theory (DMET) \cite{Knizia:2012de} is a well-known problem decomposition technique in QCC to divide a molecule into multiple fragments. DMET enables us to calculate the energy of a large-scale molecule on insufficient computational resources to handle the entire molecule by dividing the molecule into smaller fragments. Moreover, it can be combined with VQE to mitigate the noise impact of NISQ computers by decreasing the size of a quantum circuit per fragment compared to that for an entire molecule \cite{Kawashima:2021op,Shang:2022la}.

In this work, we aim to calculate the ground-state energy of the H4-chain molecule more accurately than a gold standard classical method called {\it CCSD(T)}. Our preliminary evaluation using a quantum computer simulator shows that VQE achieves a higher accuracy than CCSD(T) in absence of noises, but not with noises even if DMET is combined with VQE. Therefore, we present three approaches to mitigate the noise impact on the DMET+VQE combination. The first is {\it bath-reduced DMET} that decreases the size of quantum circuits used in VQE by reducing the number of bath orbitals which represent interactions between fragments in DMET. The second is {\it noise-less reduced density matrix (RDM) calculation} that avoids the noise impact on RDM calculation, which is necessary to calculate a molecular energy in DMET, by obtaining expectation values from quantum circuits executed using a noise-less quantum computer simulator. The third is {\it parameter refinement} that makes VQE more robust to noises by refining the parameters of quantum circuits optimized by VQE with mathematical post-processing. 

Our evaluation using a noisy quantum computer simulator shows that our three approaches enable the combination of DMET and VQE to achieve the chemical accuracy which CCSD(T) does not achieve for the H4-chain molecule. In addition, using a real NISQ device, we demonstrate that the DMET+VQE combination applying all of our three approaches achieves a higher accuracy than CCSD(T).

\section{Background}
\label{sec:background}

\subsection{Quantum Chemical Calculations (QCC)}
\label{subsec:qcc}

QCC are molecular simulation technologies to analyze the characteristics of molecules based on computational calculations. The potential energy of a target molecule can be calculated by solving the {\it \schr\ equation} with its Hamiltonian. While exactly solving it is not practical, accurate potential energies are required to analyze molecular characteristics accurately. In particular, the {\it chemical accuracy}, which is within 1.6E-3 hartree from an exact energy, is an important accuracy criterion to reproduce chemical experimental results.

To calculate molecular potential energies, a wide variety of classical methods with different accuracy levels and computational costs have been developed. For instance, {\it coupled-cluster single-double(-triple), CCSD(T),} is regarded as the gold standard in QCC because it can achieve a high accuracy with an acceptable computational cost for current classical computers~\cite{Ku:2019ac}. However, it is known that CCSD(T) cannot achieve a high accuracy for molecular structures with long-distance bonds due to a strong electron correlation~\cite{Kowalski:2000re}. 

\subsection{Variational Quantum Eigensolver (VQE)}
\label{subsec:vqe}

VQE is a hybrid quantum-classical eigensolver designed for NISQ computers and can calculate the ground-state energies of molecules for QCC \cite{Peruzzo:2014va,Tilly:2022th}. It approximates the ground-state energy of a target molecule by iteratively executing a parameterized quantum circuit on a quantum device and updating the parameters of the quantum circuit using a classical optimizer so that the molecular energy is minimized. This hybrid scheme enables VQE to work with a shallow quantum circuit, which can be executed on NISQ computers having a limited coherence time. VQE is expected to accurately calculate the ground-state energies of large-scale molecules in polynomial time by utilizing the quantum parallelism of NISQ computers.

The accuracy and computational cost of VQE strongly depend on an ansatz, which is a trial wave function to represent electronic states. A parameterized quantum circuit used by VQE is constructed based on an ansatz. It is known in QCC that the {\it unitary coupled-cluster single and double (UCCSD)} ansatz can achieve a high accuracy  \cite{Ku:2019ac,Gonthier:2022me}. However, a quantum circuit based on the UCCSD ansatz includes a lot of 2-qubit controlled NOT (CNOT) gates that are sensitive to the noises of quantum devices. Thus, it is challenging to achieve a high accuracy with VQE running on real NISQ devices.

\subsection{Density Matrix Embedding Theory (DMET)}
\label{subsec:dmet}

DMET is one of problem decomposition methods to study large-scale molecules \cite{Kawashima:2021op}. It divides a target molecule into multiple fragments and calculates the total energy of the entire molecule by calculating the energies of all fragments and summing them up. The energy of each fragment can be calculated by either a classical method or VQE. 

\begin{algorithm}[t]
    \caption{The pseudo code of DMET+VQE combination}
    \label{algo:dmet_vqe}
    \begin{algorithmic}[1]
        \REQUIRE Target molecule
        \STATE Calculate the mean field
        \STATE Divide the target molecule into fragments
        \WHILE {$\sum N_{elec,i} \neq N_{elec,mol}$}
        \STATE Determine chemical potential
            \FOR {$i \in$ fragments}
                \STATE Construct bath orbitals
                \STATE Construct embedded Hamiltonian
                \STATE Optimize parameters of ansatz with VQE
                \STATE Calculate 1- and 2-RDM with expectation values
                \STATE Calculate $N_{elec,i}$
            \ENDFOR
            \STATE Calculate total energy
        \ENDWHILE
    \end{algorithmic}
\end{algorithm}

Algorithm \ref{algo:dmet_vqe} is the pseudo code of the combination of DMET and VQE. DMET divides a target molecule into fragments based on the mean field and then repeats {\it DMET cycles} until the total number of electrons across all fragments matches that of the entire molecule. In every DMET cycle, each fragment is processed after a chemical potential, which is a variational parameter optimized across DMET cycles, is determined. For each fragment, bath orbitals which represent interactions between fragments are constructed, and the embedded Hamiltonian is constructed based on them. Then, the parameters of an ansatz-based parameterized quantum circuit is optimized with VQE for the embedded Hamiltonian. After that, one-particle and two-particle reduced density matrices (1-RDM and 2-RDM) are calculated based on the expectation values obtained by executing the quantum circuit with the optimized parameters. Finally, the number of electrons of each fragment is calculated. Once 1-RDMs and 2-RDMs of all fragments are calculated, the total energy of the target molecule is calculated based on them. 

\section{Motivation}
\label{sec:motivation}

In this work, we target the H4-chain molecule with the STO-3G basis set and aim to achieve a higher accuracy than CCSD(T) with VQE executed on a real NISQ device. However, it is challenging due to the significant impact of noise errors of current NISQ devices. 

\begin{figure}[t]
    \centering
    \subfloat{\includegraphics[width=0.4\linewidth]{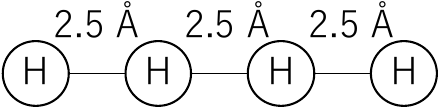}}
    \\
    \subfloat{\includegraphics[width=\linewidth]{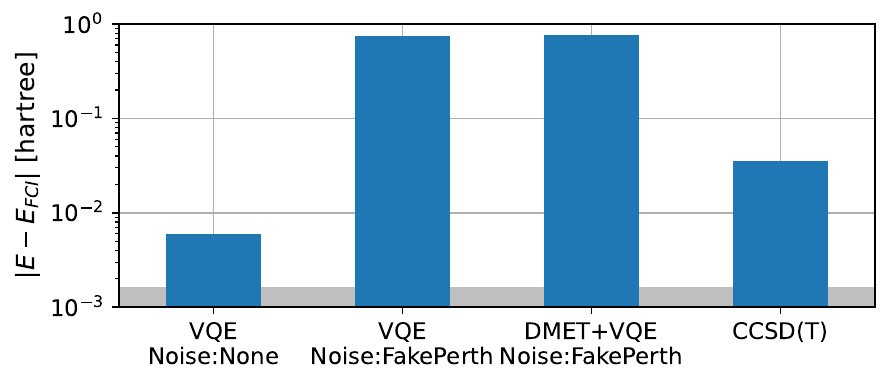}}
\caption{The energy error of VQE, naive DMET+VQE combination in the noise-less and noisy simulation and CCSD(T). The ansatz is UCCSD. The optimizers are SLSQP and SPSA in the noise-less and nosy simulation, respectively. DMET divides the H4-chain molecule into two H2 fragments. The dark shade represents the chemical accuracy.}
    \label{fig:motivation}
\end{figure}

Using a quantum computer simulator, {\it AerSimulator}, implemented in Qiskit \cite{Qiskit}, we investigate the noise impact on the accuracy of VQE for the H4-chain molecule with the bond distance of 2.5~\ang. For the noisy simulation, we use the {\it FakePerth} noise model based on the 7-qubit Perth device in IBM Quantum. The exact energy of the H4-chain molecule is obtained with the full configuration interaction (FCI) method implemented in PySCF \cite{pyscf}. \reffig{fig:motivation} plots the energy error from the exact energy, $|E-E_{FCI}|$, as an accuracy metric. The left bar shows that VQE with the UCCSD ansatz achieves a lower error than CCSD(T) in the noise-less simulation. In contrast, the middle bar shows that the error of VQE is drastically increased in the noisy simulation, leading to a higher error than CCSD(T). The right bar shows the error of the naive combination of DMET and VQE which divides the H4-chain molecule into two H2 fragments in the noisy simulation. It does not mitigate the VQE error at all because the number of orbitals of each H2 fragment (i.e., the fragment size) is same to that of the entire H4-chain molecule. The number of orbitals of the entire H4-chain molecule is four, while each H2 fragment includes two bath orbitals in addition to the two orbitals of H2.

\section{Our Approaches}
\label{sec:approaches}

To mitigate the noise impact on the VQE accuracy, we present three approaches for the combination of DMET and VQE.

\subsection{Bath-Reduced DMET}
\label{subsec:br-dmet}

The first approach is {\it bath-reduced DMET} that reduces the number of bath orbitals in each fragment to a specified value. When DMET divides a molecule into multiple fragments, it sustains a high accuracy by  representing interactions between fragments using bath orbitals. On the other hand, the noise impact on VQE  is larger for fragments having more orbitals, because the number of orbitals in each fragment determines the number of qubits of a quantum circuit used by VQE. Thus, there is a tradeoff between including more bath orbitals to represent interactions between fragments and including less bath orbitals to reduce the number of qubit of a quantum circuit. 

Therefore, we implement a new option of DMET to adjust the number of bath orbitals in each fragment. DMET originally determines the number of bath orbitals in each fragment by scoring them based on their energies and comparing the scores with a threshold. If the number of bath orbitals is specified with the new option, the specified number of bath orbitals having higher scores are only included. 

\reftab{tab:example_bath_reduced_dmet} shows an example of bath-reduced DMET for the H4-chain molecule. When DMET divides the H4-chain molecule into two H2 fragments, it originally includes two bath orbitals in each H2 fragment. In this case, the total number of orbitals in each fragment is four (two of H2 plus two bath orbitals), and the number of qubits of a quantum circuit for each fragment is eight. Note that the number of qubits is twice of the number of orbitals with consideration of spin orbitals. If we reduce the number of bath orbitals to one and zero, the number of qubits is reduced to six and four, respectively. Furthermore, a well-known qubit tapering method called {\it Z2 symmetry reduction}~\cite{bravyi:2017ta} can reduce the numbers of qubits to six, four, and two when the numbers of bath orbitals are two, one, and zero, respectively.

\begin{table}[t]
    \caption{An example of bath-reduced DMET for the H4-chain molecule.}
    \begin{center}
    \begin{tabular}{c|ccc}
    \hline
    \# of bath & Total \# of & \multirow{2}{*}{\# of qubits} & \# of qubits \\
    orbitals & orbitals &  & w/ Z2 symm. red. \\
    \hline
    2 (original) & 4 & 8 & 5 \\
    1 & 3 & 6 & 4 \\
    0 & 2 & 4 & 2 \\
    \hline
    \end{tabular}
    \label{tab:example_bath_reduced_dmet}
    \end{center}
    \end{table}

\subsection{Noise-Less RDM Calculation}
\label{subsec:nlrdm}

The second approach is {\it noise-less RDM calculation} that calculates 1-RDM and 2-RDM of each fragment based on expectation values obtained from a quantum circuit executed in the noise-less simulation. As shown in Algorithm \ref{algo:dmet_vqe}, VQE optimizes the parameters of an ansatz-based quantum circuit (line 8), and then 1-RDM and 2-RDM are calculated based on expectation values obtained by executing the quantum circuit with the optimal parameters (line 9). VQE is robust to noises in some extent but time-consuming due to the iterative quantum circuit executions. In contrast, expectation values for RDM calculation are obtained only once, which is sensitive to noises.

Therefore, we use a real NISQ device to optimize the parameters of an ansatz-based quantum circuit with VQE (line 8) and use a noise-less quantum computer simulator to obtain expectation values for RDM calculation (line 9). This approach can avoid the noise impact on the noise-sensitive RDM calculation. Note that it requires a quantum computer simulator supporting the same number of qubits to that of the quantum circuit.

\subsection{Parameter Refinement}
\label{subsec:parameter_refinement}
The third approach is {\it parameter refinement} that is a post-processing method to refine the parameters optimized by a classical optimizer in the VQE optimization process. Using Gaussian process (GP), 
this method takes a history of parameter values as an input and returns the ``refined'' parameters of an ansatz which are robust to noises. 

\begin{figure}[t]
    \includegraphics[width=0.5\textwidth]{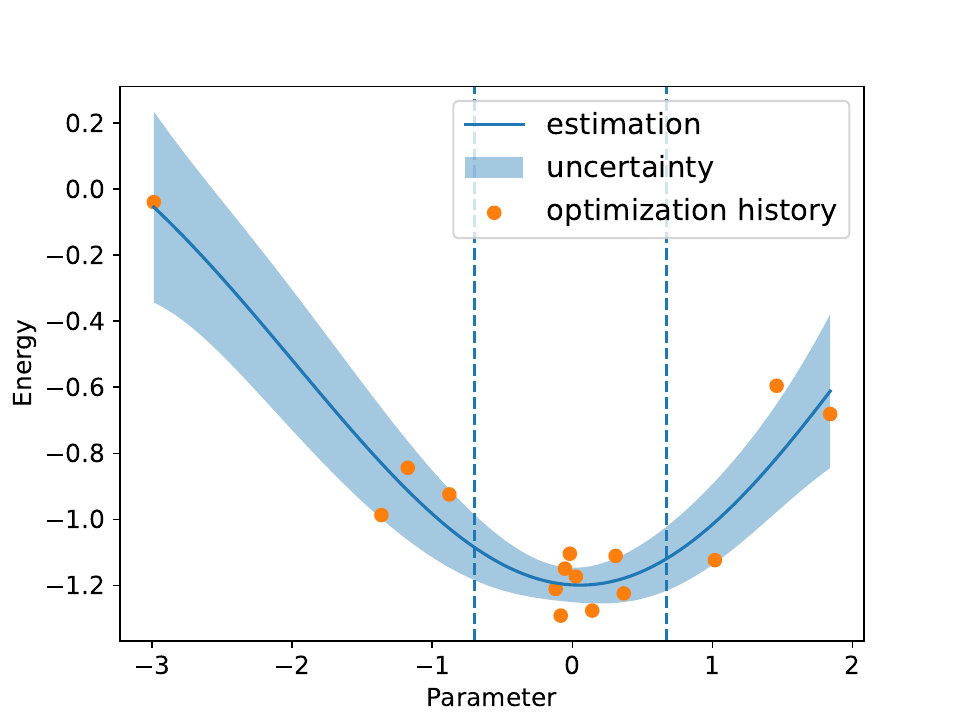} 
    \caption{An example of parameter refinement using Gaussian process. The orange points represent energies observed by an optimizer with varied parameters. The blue curve represents the estimation using the observation, and the blue shade represents uncertainty in the estimation.}
    \label{fig:parameter-refinement-explanation}
\end{figure}

\reffig{fig:parameter-refinement-explanation} shows how the parameter refinement method finds the refined parameter. It takes a history of an energy and parameter (orange dots) obtained from an optimizer such as SPSA, and construct the estimation of the energies (the blue curve). Then, it finds the refined parameter that minimizes the estimated energy in the domain where uncertainty is lower than a threshold (the domain between the vertical doted lines).

We explain the method in detail with the brief review of an online estimator using the GP regression~\cite{rasmussen2006gaussian,wendland2004scattered,srinivas2010gaussian,chowdhury2017kernelized,durand2018streaming}.
We denote by $\Theta \subset \RR^d$ the space of parameters (e.g., $\Theta = [-\pi, \pi]^d$),
where $d$ is the dimension of the parameter.
Let $k: \Theta \times \Theta \rightarrow \RR$ be a semi-positive definite kernel. 
The kernel $k(\theta, \theta')$ represents the similarity of two parameters $\theta$ 
and $\theta'$.
There are some types of kernels such as Mat\'ern-$\nu$ kernel with the smoothness parameter $\nu$~\cite{van2011information} and the squared exponential kernel defined as $k(\theta, \theta') = \exp(-\|\theta - \theta'\|^2/l)$, where $l > 0$ is the length-scale parameter. 

Next, we define the estimation $\mu_n(\theta)$ and uncertainty $\sigma_n(\theta)$ using GP regression.
Let $(\theta_1, y_1), \dots, (\theta_n, y_n)$ be a history of a parameter and energy obtained in the optimization process in VQE. More precisely, $y_i$ is the energy obtained from an ansatz with the parameter $\theta_i$ which has been updated by an optimizer such as SPSA. Let $\lambda > 0$ be a parameter of the estimation called regularizer. For $\theta \in \Theta$, we define $\mu_n(\theta) \in \RR$ and $\sigma_n(\theta) \in \RR_{\ge 0}$ as follows:
\begin{align}
    \mu_n(\theta) &= (y_1,\dots, y_n) (K_n + \lambda 1_n)^{-1} k_n(\theta), \label{eq:gp-mu}\\
    \sigma_n^2(\theta) &= k(\theta, \theta) - k_n(\theta)^\top (K_n + \lambda 1_n)^{-1} k_n(\theta).\label{eq:gp-sigma}
\end{align}
Here, $1_n\in \RR^{d\times d}$ is the identity matrix, $K_n \in \RR^{n \times n}$ is a symmetric matrix such that the $ij$-entry of $K_n$ is given as $k(\theta_i, \theta_j)$, and $k_n(\theta) \in \RR^{n}$ is a column vector defined as $(k_n(\theta))_i = k(\theta, \theta_i)$. Although the computational complexity of these quantities is $O(n^3)$, it can be converted to $O(poly(d))$ under a natural assumption. Since the typical convergence rate of an optimization algorithm using estimated gradients is $O(poly(d)/\sqrt{t})$, where $t$ is the number of total iterations (see \cite[Chapter 6]{bubeck2015convex} and \cite{agarwal2010optimal}), it is natural to assume $n=O(poly(d))$. Moreover, if $n$ is large, kernel approximation methods can be applied to reduce the computational complexity \cite{santin2017convergence,mutny2018efficient}.

Let $f:\Theta \rightarrow \RR$ be a function that returns the noiseless energy $f(\theta)$.
If samples are generated by $y_i = f(\theta_i) + \varepsilon_i$, where $\{\varepsilon_i\}_{1\le i \le n}$ 
are independent random variable that follows the normal distribution $\mathcal{N}(0, R^2)$,
then under mild assumptions on $f$, 
for any $\theta \in \Theta$, we have the following inequality with probability at least $1-\delta$:
\begin{equation}
    \left|f(\theta) - \mu_n(\theta) \right| \le 
    \sigma_n(\theta) \left(
        \|f \| + \frac{R}{\lambda} \sqrt{2 \log (1/\delta) + 2\gamma_{n, \lambda}}
    \right),
    \label{eq:rkhs-confidence-interval}
\end{equation}
where $\|f\|$ and $\gamma_{n, \lambda}$ are constants independent of $\theta$.
We refer to \cite[Theorem 1]{durand2018streaming} for the mathematically precise statement.
By this inequality, we see that $\mu_n(\theta)$ represents an estimation of the noise-less energy $f(\theta)$, and $\sigma_n(\theta)$ represents uncertainty in the estimation (i.e., the width of the confidence interval of $f(\theta)$).
We also note that there is a tradeoff of the choice of $\lambda$; if $\lambda$ is too small, then $\mu_n(\theta)$ overfits to noisy data and the confidence interval becomes wider due to the second term in the RHS in \eqref{eq:rkhs-confidence-interval}. Ideally, $\lambda$ should be selected at the same order as the noise level $R$ \cite{durand2018streaming}.

Algorithm~\ref{algo:parameter-refinement} shows the pseudo code of our parameter refinement method. It takes a history of the VQE optimization $(\theta_1, y_1), \dots, (\theta_n, y_n)$, kernel $k$, regularizer $\lambda$, and $c>0$, and returns a refined parameter $\theta^*$. We consider the optimization objective \eqref{eq:pr-obj} to minimize the surrogate function $\mu_n(\theta)$ of the VQE energy subject to the condition that uncertainty $\sigma_n(\theta)$ of the estimation is smaller than the given threshold $c$.

Algorithm \ref{algo:parameter-refinement} has some similarity to Bayesian optimization algorithms \cite{srinivas2010gaussian,iannelli2022noisy} and methods using Gaussian Process as a surrogate model \cite{muller2022accelerating}. However, our parameter refinement method is post-processing applied only once after the VQE optimization, while the methods in \cite{iannelli2022noisy,muller2022accelerating} repeatedly optimize a surrogate model in the optimization process. Thus, our method can be combined with any optimization algorithms such as SPSA and NFT \cite{nakanishi2020sequential}.

\begin{algorithm}[t]
    \caption{The pseudo code of parameter refinement}
    \label{algo:parameter-refinement}
    \begin{algorithmic}[1]
        \REQUIRE kernel $k$, regularizer $\lambda$, a history of a parameter and energy $(\theta_1, y_1), \dots, (\theta_n, y_n)$, a search domain of parameters $\Theta' \subset \Theta$, and a threshold $c > 0$.
        \STATE Optimize parameters of the kernel using the data (see \cite[Chapter 5.4]{rasmussen2006gaussian})
        \STATE Solve the following optimization problem:
        \begin{equation}
            \label{eq:pr-obj}
            \min_{\theta \in \Theta'} \mu_n(\theta) \text{ subject to } \sigma_n(\theta) \le c.
        \end{equation}
        \STATE Return the solution of $\theta^*$ of Eq. \eqref{eq:pr-obj}.
    \end{algorithmic}
\end{algorithm}

\section{Evaluation}
\label{sec:evaluation}

In this section, we evaluate the accuracy of the combination of DMET and VQE with our three approaches using a quantum computer simulator and a real NISQ device. We first explain our experimental setup and then show the evaluation results.

\subsection{Experimental Setup}
\label{subsec:setup}

We implement VQE with {\it Qiskit} \cite{Qiskit} and use {\it AerSimulator} for the noise-less and noisy simulation. For the noisy simulation, we use the {\it FakePerth} noise model, which is based on the calibration data of the 7-qubit {\it Perth} quantum device in IBM Quantum. VQE uses the UCCSD ansatz, the SLSQP optimizer in the noise-less simulation, and the SPSA optimizer in the noisy simulation. The UCCSD-based quantum circuits are transpiled with the optimization level 3 and executed with 1,000 shots. 

The target molecule is H4-chain with the STO-3G basis set, which is a chain-structured molecule with four hydrogen atoms. The Hamiltonian is mapped to qubits using the Jordan-Wigner mapper~\cite{Fradkin:1989jo}, and the number of qubits is tapered with the Z2 symmetry reduction method~\cite{bravyi:2017ta}. As the accuracy criterion, we obtain the exact ground-state energy of the H4-chain molecule with the full configuration interaction (FCI) method implemented in {\it PySCF} \cite{pyscf}.

We run DMET to divide the H4-chain molecule into two H2 fragments with Tangelo \cite{Tangelo}, which is an open-source Python package for material simulation on quantum computers. Although Tangelo includes a unique VQE implementation, we add the VQE implemented with Qiskit as a new solver because it is faster. We set the maximum DMET cycles to ten and report the minimum energy among all cycles as the output energy. For noise-less RDM calculation, we obtain expectation values by executing the UCCSD-based quantum circuit with the optimal parameters in the noise-less simulation. For parameter refinement, we use the Mat\'ern-$\nu$ kernel with the smoothness parameter $\nu$ of 5.5, the regularizer $\lambda$ of 1e-4, and the threshold $c$ of 1.0.

\subsection{Benefit of Bath-Reduced DMET}

\begin{figure}[t]
    \centering
    \subfloat{\includegraphics[width=0.4\linewidth]{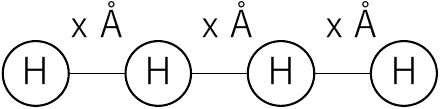}}
    \\
    \setcounter{subfigure}{0}
    \subfloat[Noise-less simulation]{
        \includegraphics[width=0.93\linewidth]{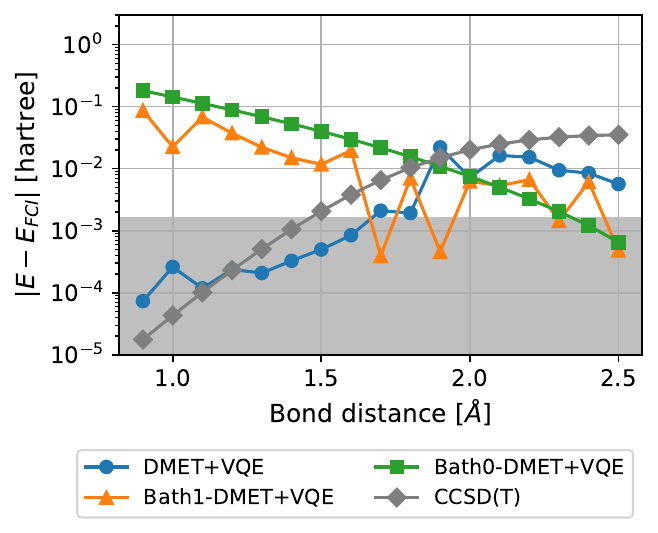}
        \label{subfig:pec_noiseless_simulation}}
    \\
    \subfloat[Noisy simulation with the FakePerth noise model]{
        \includegraphics[width=0.93\linewidth]{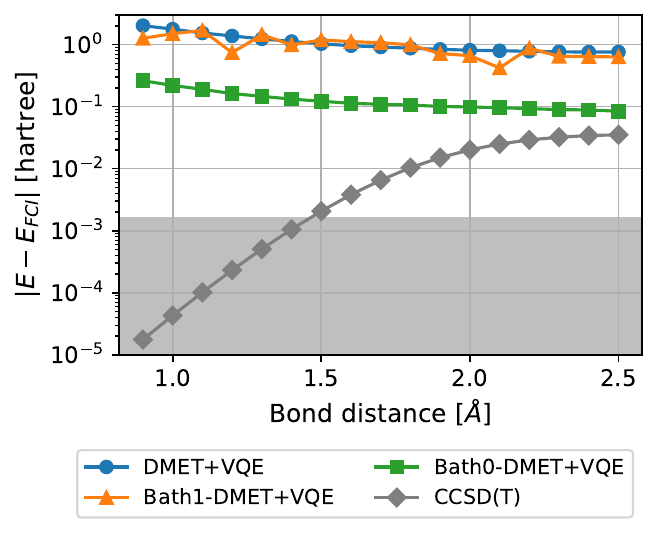}
        \label{subfig:pec_noisy_simulation}}
    \caption{The energy errors of four methods along with bond distances. The dark shade represents the chemical accuracy.}
    \label{fig:bath_reduced_dmet}
\end{figure}

We first evaluate the benefit of bath-reduced DMET. \reffig{subfig:pec_noiseless_simulation} plots the energy errors of four different methods along with different bond distances in the noise-less simulation. The naive combination of DMET and VQE shown with the blue line satisfies the chemical accuracy at bond distances of 1.6~\ang\ or shorter by accurately capturing interactions between the two fragments with two bath orbitals per fragment. Moreover, it achieves lower errors than CCSD(T) at bond distances of 1.3\ang\ or longer. In contrast, Bath1-DMET+VQE and Bath0-DMET+VQE (orange and green lines), which reduce the number of bath orbitals to one and zero respectively, incur high errors at short bond distances. However, interestingly, they achieve lower errors than the naive DMET+VQE combination (blue line) and CCSD(T) (gray line) at long bond distances. This result demonstrates that bath-reduced DMET sustains a high accuracy for the H4-chain molecule having a long bond distance even if it reduces the number of bath orbitals per fragment.

\reffig{subfig:pec_noisy_simulation} shows the results in the noisy simulation. We can see in this graph that the errors of DMET+VQE (blue line) and Bath1-DMET+VQE (orange line) are especially high, whereas Bath0-DMET+VQE (green line) achieves an order of magnitude lower errors than them. We explain this reason with \reftab{tab:cnot_bath_reduced_dmet}. With DMET+VQE and Bath1-DMET+VQE, the UCCSD ansatz has hundreds of CNOT gates and depth. In contrast, Bath0-DMET+VQE reduces the number of CNOT gates to four and the depth to ten. Since CNOT gates are especially sensitive to noises, the significant reduction of CNOT gates mainly contributes to the low error of Bath0-DMET+VQE. However, Bath0-DMET+VQE (green line) does not still outperform CCSD(T) (gray line) at any bond distances, and its accuracy is still far from the chemical accuracy.

\begin{table}[t]
    \caption{The number of CNOT gates and depth of the UCCSD ansatz for the H4-chain molecule with bath-reduced DMET. The Z2 symmetry reduction is applied.}
    \begin{center}
    \begin{tabular}{c|ccc}
    \hline
     & \# of qubits & \# of CNOT gates & Depth \\
    \hline
    DMET+VQE & 5 & 336 & 455 \\
    Bath1-DMET+VQE & 4 & 172 & 235 \\
    Bath0-DMET+VQE & 2 & 4 & 10 \\
    \hline
    \end{tabular}
    \label{tab:cnot_bath_reduced_dmet}
    \end{center}
\end{table}

\subsection{Benefits of Noise-less RDM calculation and Parameter Refinement}

\begin{figure}[t]
    \centering
    \subfloat{\includegraphics[width=0.4\linewidth]{h4chain_d25.pdf}}
    \\
    \subfloat{\includegraphics[width=\linewidth]{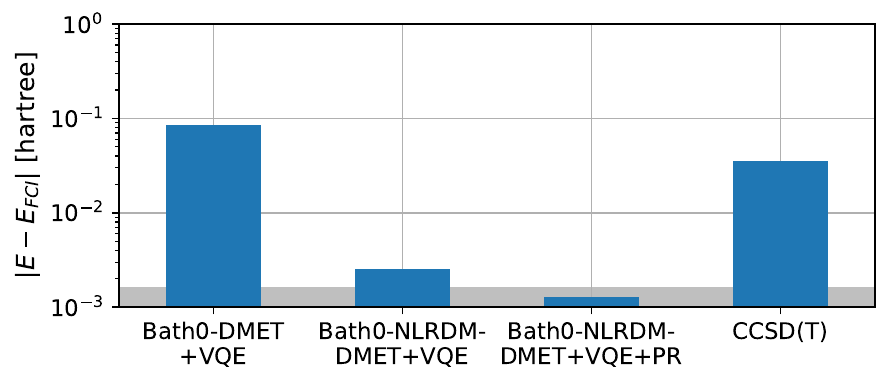}}
\caption{The energy error of the DMET+VQE combination with our approaches in the noisy simulation and CCSD(T). The dark shade represents the chemical accuracy.}
    \label{fig:nlrdm_pr_noisy_simulation}
\end{figure}

Next, we evaluate the benefits of the noise-less RDM calculation and parameter refinement. \reffig{fig:nlrdm_pr_noisy_simulation} plots the error of the DMET+VQE combination with three patterns at a bond distance of 2.5~\ang\ in the noisy simulation. The left bar shows the error of Bath0-DMET, which is identical to the rightmost point of the green line in \reffig{subfig:pec_noisy_simulation}. The middle bar shows that the noise-less RDM calculation (NLRDM) reduces the error of Bath0-DMET significantly and enables us to achieve an order of magnitude lower error than CCSD(T). This result demonstrates that we successfully avoid the noise impact on the RDM calculation by using the noise-less quantum computer simulator. In addition, the right bar shows that parameter refinement (PR) further reduces the error, and applying all of our three approaches achieves the chemical accuracy in the noisy simulation.

\subsection{Evaluation using a Real NISQ device}

\begin{figure}[t]
    \centering
    \subfloat{\includegraphics[width=0.4\linewidth]{h4chain_d25.pdf}}
    \\
    \subfloat{\includegraphics[width=\linewidth]{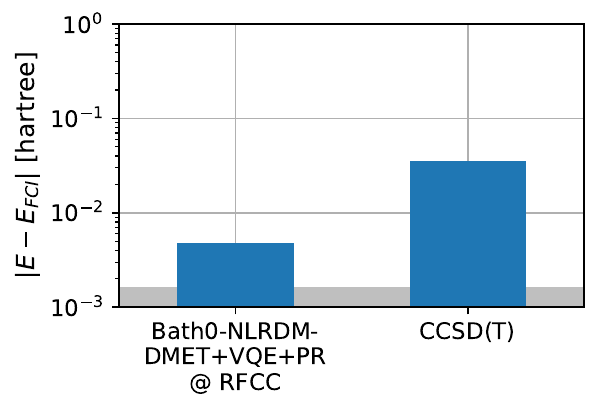}}
\caption{The energy error of the DMET+VQE combination applying all of our three approaches on real NISQ devices and CCSD(T). The dark shade represents the chemical accuracy.}
    \label{fig:error_real_nisq_devices}
\end{figure}

Finally, we evaluate the accuracy of the DMET+VQE combination using a real 64-qubit NISQ device in RIKEN RQC - FUJITSU Collaboration Center (RFCC). To mitigate the noise impact of them as much as possible, we apply Bath0-DMET, noise-less RDM calculation, and parameter refinement. Thus, two qubits are used on the device. The 1-qubit gate fidelity measured via randomized benchmarking \cite{randomized_bench} before our evaluation is 0.998 on both the qubits, and the 2-qubit gate fidelity is 0.97. In addition, a readout mitigation method called {\it measurement calibration}~\cite{measurement_calibration} is applied. We reduce the number of quantum circuits executed in VQE to 4 by grouping fragment's hamiltonian consisted of 8 Pauli strings into four groups as shown in \reftab{tab:grouping} compared to 8 circuits for individual measurement. We use the UCCSD ansatz and SPSA optimizer in similar to the noisy simulation. The UCCSD-based quantum circuits built with Qiskit are transpiled to quantum circuits composed of basis gates supported by the device. The number of shots is set to 10,000.
\reffig{fig:error_real_nisq_devices} shows that the DMET+VQE combination applying all of our three approaches achieves much lower errors than CCSD(T) on the real NISQ device.

\begin{table}[t]
    \caption{Hamiltonian grouping}
    \begin{center}
    \begin{tabular}{c|r}
    \hline
    Group & Pauli strings \\
    \hline
    1 & IZ, ZI, ZZ \\
    2 & IX, ZX \\
    3 & XI, XZ \\
    4 & XX \\
    \hline
    \end{tabular}
    \label{tab:grouping}
    \end{center}
\end{table}

\section{Conclusion}
In this work, we present three approaches to mitigate the noise impact of NISQ computers on the combination of DMET and VQE: bath-reduced DMET, noise-less RDM calculation, and parameter refinement. For the H4-chain molecule, the evaluation in the noisy simulation shows that the three approaches significantly improve the accuracy of the DMET+VQE combination and enable us to achieve the chemical accuracy. In addition, the evaluation using a real NISQ device demonstrates that the DMET+VQE combination applying all of the three approaches achieves a higher accuracy than a gold standard classical method, CCSD(T). As our future work, we will evaluate the accuracy of the DMET+VQE combination with larger various molecules.

\section*{Acknowledgements}
We are grateful to the RIKEN Center for Quantum Computing for the preliminary use of the RIKEN RQC quantum computer developed under the support by MEXT Q-LEAP (Grant No. JPMXS0118068682). We also would like to thank all of our colleagues in the Quantum Laboratory, Fujitsu Research. Their great effort helped us obtain the evaluation results in this work. 

\bibliographystyle{IEEEtran}
\bibliography{references}

\appendix
\section*{Selection of Regularizer for the Parameter Refinement Method}
In this section, we introduce methods for selecting 
the regularizer $\lambda$ that is an input of Algorithm \ref{algo:parameter-refinement}.
As explained in Section \ref{subsec:parameter_refinement}, the regularizer $\lambda$ 
and should be selected as the same order of the noise level of observations.
If additional measurements of Ansatz are allowed, then by evaluating the returned parameter 
$\theta^*_{\lambda}$ by Algorithm \ref{algo:parameter-refinement}
using multiple choices of $\lambda \in L$, one can select the regularizer in a predefined finite set $L$
of regularizers.
If additional measurements are not allowed and we do not have a prior knowledge of the noise level,
then one can use a standard method to select a regularizer, i.e., evaluating $\mu_n(\theta)$ on a validation dataset
for multiple choices of $\lambda$.
However, there are differences than the standard setting of the supervised learning and more challenging 
due to the following reasons:
(i) the dataset $(\theta_1, y_1), \dots, (\theta_n, y_n)$ is not i.i.d generated and samples are highly correlated
since it is generated by an optimization algorithm
and (ii) the objective of the estimation is not to minimize the error between the estimation and the noiseless energy in the entire domain, but 
to minimize the error of the estimation on a neighborhood of the optimal point. 
Therefore, to avoid overfitting to noisy samples, 
one has to take a validation dataset as a set of samples $(\theta_i, y_i)$ belonging to
a neighborhood of the optimal point (e.g., a neighborhood of the solution returned by the optimizer).
We introduce an alternative method to select a regularizer $\lambda$ in a finite set $L$ of candidates.
Here, following \cite{nakanishi2020sequential},
we assume that the noiseless energy $f(\theta)$ associated $\theta$ is a sin function 
$a \sin(\theta(i) + \theta_0) + b$ 
of the $i$-th entry $\theta(i)$ of $\theta$ and fix the other entries,
where $a, b, \theta_0$ are constants independent of $\theta(i)$.
We denote by $\widetilde{\theta}$ the solution returned by the optimizer (e.g. SPSA)
and denote by $\theta(j)$ the $j$-th entry of $\theta$.
To select $\lambda \in L$, we fix the index $i$ of the entry and generate
 a dataset $(\theta'_1, y_1'), \dots, (\theta'_m, y_m')$, where 
$\theta'_l(j) = \widetilde{\theta}(j)$ for all $i \neq j$ and $1 \le l \le m$,
$\theta'_l(i)$ is uniform randomly generated around $\widetilde{\theta}(i)$,
and $y_l' = \mu_{n, \lambda}(\theta_l')$,
where $\mu_{n, \lambda}$ denotes $\mu_n$ using the parameter $\lambda$.
Then, using the dataset $(\theta'_1, y_1'), \dots, (\theta_m', y_m')$,
we first fit parameters $a, b, \theta_0$ of the function
$\theta \mapsto a\sin(\theta(i) + \theta_0) + b$ using a loss function.
We generate a similar but independent dataset $(\theta'_1, y_1'), \dots, (\theta'_m, y_m')$
and compute a validation error $e_\lambda$ using the independent dataset and the learned function
$a\sin(\theta(i) + \theta_0) + b$.
Then, we select $\lambda \in L$ that gives the minimum validation error $e_\lambda$.
\end{document}